\documentclass[10pt,conference]{IEEEtran}
\IEEEoverridecommandlockouts
\usepackage[utf8]{inputenc}
\usepackage{wrapfig}
\usepackage{graphicx}
\usepackage{pgfplots}
\usepackage{pgfplotstable}
\usepackage{tcolorbox,fancyvrb,xcolor,tikz}
\usepackage{multirow}
\usepackage{fancyvrb}

\usepackage{cite}
\usepackage{hyperref}

\usepackage{algorithmic}
\usepackage{textcomp}
\usepackage{xcolor}
\usepackage{booktabs}
\usepackage{siunitx}

\usepackage{color, colortbl}

\usepackage{xspace}

\usepackage{listings}

\usepackage{flushend} 

\colorlet{punct}{red!60!black}
\definecolor{background}{HTML}{EEEEEE}
\definecolor{delim}{RGB}{20,105,176}
\colorlet{numb}{magenta!60!black}

\def\BibTeX{{\rm B\kern-.05em{\sc i\kern-.025em b}\kern-.08em
    T\kern-.1667em\lower.7ex\hbox{E}\kern-.125emX}}

\begin{document}

\title{Technical Debt Friction for Maintenance Prioritization: An Industrial Multi-Case Study}

%\author{arda.goknil }
%\date{December 2021}

%ARDA: Alternative titles:

%\title{Is Technical Debt Friction Actionable? An Industrial Validation with CodeScene-Based Analysis}

%\title{Validating Technical Debt Friction in Industrial Practice: A Multi-Case Study}

%\title{Is Technical Debt Friction Actionable? Evidence from Multiple Industrial Cases}

%\title{Technical Debt Friction for Maintenance Prioritization: An Industrial Multi-Case Study}

%\title{Understanding Technical Debt Friction in Practice: Evidence from Industrial Software Systems}

%\title{From Code Health to Maintenance Friction: A Multi-Case Industrial Study}

%\title{Validating Technical Debt Friction in Industrial Practice: Insights from Discussions with Software Practitioners}

%\title{From Code Health to Friction: An Industrial Perspective on Prioritizing Maintenance Work}

%\title{Understanding Technical Debt Friction in Practice: An Exploratory Industry Study}
%\title{Industrial Validation of Technical Debt Friction for Maintenance Decision Support}

\author{\IEEEauthorblockN{Simeon Tverdal\IEEEauthorrefmark{1}, Phu Nguyen\IEEEauthorrefmark{1}, Arda Goknil\IEEEauthorrefmark{1}, Antonio Martini\IEEEauthorrefmark{2}, \\Markus Borg\IEEEauthorrefmark{3}\IEEEauthorrefmark{4}, and Adam Tornhill\IEEEauthorrefmark{3}  %XXX YYY\IEEEauthorrefmark{4}
}

\IEEEauthorblockA{\IEEEauthorrefmark{1}\textit{SINTEF}, 
%\textit{Sustainable Communication Technologies}\\
Oslo, Norway, \{firstname.lastname\}@sintef.no}

\IEEEauthorblockA{\IEEEauthorrefmark{2}\textit{University of Oslo}, Oslo, Norway,
\{antonima\}@ifi.uio.no}

\IEEEauthorblockA{\IEEEauthorrefmark{3}\textit{CodeScene}, Malmö, Sweden
\{firstname.lastname\}@codescene.com}

\IEEEauthorblockA{\IEEEauthorrefmark{4}\textit{Lund University}, Lund, Sweden}

% \IEEEauthorblockA{\IEEEauthorrefmark{4}\textit{AKVA Group}, Klepp, Norway,
% \{?\}@akvagroup.com}

}

\maketitle

\begin{abstract}

Software-intensive organizations need effective ways to identify where maintenance and refactoring efforts will yield the greatest practical benefit. Although software analytics such as code health, hotspots, and coupling provide valuable signals, they do not always capture the experienced burden of change that slows software evolution in practice. This paper presents a multi-case industrial study of \emph{technical debt friction} as a prioritization-oriented concept for identifying where technical debt most strongly affects maintenance and evolution. We investigate how practitioners interpret the concept, whether friction-related analysis aligns with perceived maintenance pain points and refactoring needs, and what broader maintenance and evolution insights friction can provide beyond individual refactoring candidates. To this end, we conducted structured walkthrough sessions with practitioners across multiple industrial cases using analysis artifacts including code health, hotspots, coupling, refactoring targets, and socio-technical views. Our findings show that practitioners generally considered technical debt friction useful for reasoning about maintenance burden, especially when interpreted together with complementary technical and socio-technical views. At the file level, friction often aligned with known problematic areas and, in several cases, with files that later received maintenance attention, although its practical relevance depended strongly on context. In addition, our exploratory project-level analysis suggests that friction distributions may reveal broader maintenance and evolution patterns. These results indicate that technical debt friction is promising as a decision-support concept, but most effective when used with contextual knowledge and supporting evidence.

\end{abstract}

\maketitle

\section{Introduction}
\label{sec:introduction}

Software-intensive organizations continuously face the challenge of evolving complex systems under limited time and resource constraints~\cite{de2008empirical}. In such settings, maintenance and evolution decisions are rarely driven by the goal of improving all problematic code equally; instead, teams must decide where intervention is most worthwhile and which parts of a system most urgently deserve refactoring or restructuring~\cite{kim2012field, nagappan2005use}. Existing software analytics provide useful signals for this purpose, including code health indicators, hotspots, change coupling, and ownership-related metrics~\cite{nagappan2005use, zimmermann2005mining, bird2011don, tverdal2025combining}. However, in practice, teams need a more direct way to reason about the \emph{experienced burden of change}: not only whether code appears problematic in isolation, but whether it %actively 
slows down development, complicates maintenance, increases coordination effort, or raises the cost of future evolution~\cite{cataldo2008socio, bird2011don}.

This practical concern motivates the notion of \emph{technical debt friction}. In this paper, we use the term technical debt friction to denote the practical resistance that problematic parts of a system impose on ongoing maintenance and evolution~\cite{CodeSceneTechnicalDebtFriction}. To the best of our knowledge, technical debt friction is not a standard term in the technical debt literature; rather, it appears to have emerged as a practice-oriented operationalization of how technical debt affects development productivity and change effort~\cite{CodeSceneTechnicalDebtFriction, avgeriou2015reducing}. Rather than focusing narrowly on structural quality defects, technical debt friction aims to capture the resistance that parts of a system create during ongoing change and maintenance. Such resistance may arise from poor code quality, but also from broader factors such as recurring change concentration, unstable dependencies, refactoring difficulty, fragmented ownership, or knowledge bottlenecks~\cite{zimmermann2005mining, de2008empirical, kim2012field, bird2011don, cataldo2008socio, astekin2025detecting}. For industrial teams, this perspective is attractive because it speaks directly to day-to-day engineering concerns: which parts of the system are most difficult to work with, which areas consume disproportionate effort, and where improvement work is most likely to pay off.

Despite this promise, the usefulness of technical debt friction cannot be assumed. A concept intended for industrial decision making must be understandable to practitioners, aligned with their experience, and sufficiently actionable to support prioritization. Otherwise, it risks becoming yet another abstract software quality label with limited practical value. This creates an important problem for both researchers and tool builders: while many software analytics concepts are technically plausible, far fewer have been examined in terms of how practitioners actually interpret them, whether they trust them, and how they use them when reasoning about real maintenance pain points. In particular, it remains unclear whether technical debt friction provides added value beyond more established views such as code health, hotspots, or coupling analyses, or whether practitioners instead see it as a relabeling of existing quality concerns.

To address this problem, we conducted a \emph{multi-case industrial study} to investigate the practical usefulness of technical debt friction in software maintenance and evolution. Our study examines how practitioners interpret the concept when applied to their own systems, whether friction-related signals align with their perceptions of maintenance pain points and refactoring needs, and what benefits and limitations emerge when the concept is discussed in real industrial contexts. %More specifically, 
The study addresses the following research questions (RQs):

\begin{itemize}
    \item \textbf{RQ1:} How do practitioners interpret technical debt friction when applied to real industrial software systems?
    \item \textbf{RQ2:} To what extent does technical debt friction align with practitioners' perceptions of maintenance pain points and refactoring needs?
    \item \textbf{RQ3:} Can project-level friction distributions provide broader maintenance and evolution insights beyond individual refactoring candidates?
    %%%%\item \textbf{RQ3:} What broader maintenance and evolution insights could Technical Debt Friction provide beyond individual refactoring candidates?
    %%%\item \textbf{RQ3:} What benefits, limitations, and adoption challenges emerge when using technical debt friction for maintenance prioritization in practice?
\end{itemize}

%ARDA
%%%%The study is based on multiple industrial cases and involves practitioners with direct knowledge of the systems under analysis. Using structured walkthrough sessions centered on analysis artifacts obtained using CodeScene SE intelligence platform~\cite{codescene}, such as code health, hotspots, coupling, refactoring targets, and socio-technical views, we collected evidence on the interpretability, credibility, and actionability of the friction concept across cases.

The study covers multiple industrial cases and involves practitioners with direct knowledge of the analyzed systems. Using structured walkthrough sessions centered on CodeScene~\cite{codescene} analysis artifacts, including code health, hotspots, coupling, refactoring targets, and socio-technical views, we collected evidence on the interpretability, credibility, and actionability of the friction concept across cases.

%ARDA
%%%%%Our findings show that technical debt friction is perceived by practitioners as a meaningful way of reasoning about maintenance burden, especially when grounded in concrete examples from real systems. The concept appears most useful when it integrates multiple technical and socio-technical signals rather than being treated as a standalone metric. Across cases, practitioners related friction not only to problematic code, but also to recurring change effort, coordination overhead, ownership concentration, and areas that repeatedly complicate modification. At the same time, our results show that the concept requires careful operationalization and explanation: its value depends on how clearly it is distinguished from adjacent notions such as code health or technical debt more broadly, and on how transparently its underlying signals can be interpreted. These findings suggest that technical debt friction is promising as an industry-facing concept for maintenance prioritization, while also highlighting the conditions under which it can be applied effectively in practice.

Our findings show that practitioners generally regarded \emph{technical debt friction} as a meaningful and useful concept for reasoning about maintenance burden, but not as a standalone signal. Across the studied products, friction was most useful when discussed together with complementary views such as code health, hotspots, coupling, and socio-technical information, and when interpreted in relation to product context, architectural centrality, and ongoing development activity. At file level, friction often aligned with areas that practitioners already considered problematic and, in several cases, with files that later received maintenance or refactoring attention. At the same time, the study also showed that high friction does not uniformly imply high practical importance: isolated files, test code, and temporary activity spikes required more careful interpretation. Beyond individual candidates, our exploratory project-level analysis further suggests that friction distributions may provide broader insights into how maintenance effort is distributed across a system. These findings indicate that technical debt friction is promising as a maintenance-oriented decision-support concept, but that its practical value depends on contextual interpretation and supporting evidence.

%ARDA
%%%%%In summary, this paper makes three contributions. First, it presents a multi-case industrial study of technical debt friction in real software development contexts. Second, it provides empirical findings on how practitioners interpret friction and how friction-related analyses align with maintenance-relevant areas in practice. Third, it extends the analysis from individual files to project-level friction distributions, showing how friction can support broader reasoning about maintenance and evolution.

%%%%%The remainder of the paper is structured as follows. Section~X introduces the background and conceptual framing of technical debt friction. Section~X describes the study design, cases, participants, and analysis procedure. Section~X presents the results. Section~X discusses cross-case implications for research and practice. Section~X concludes the paper.

The paper is structured as follows. Section~\ref{sec:background} presents the background. %regarding TD and detection tools. 
In Sections~\ref{sec:research-design} and~\ref{sec:results}, we present our research design and the results, respectively. In Section~\ref{sec:discussion}, we discuss cross-case findings. Section~\ref{sec:empirical:validity} summarizes the threats to the validity of our study. Section~\ref{section:related-work} discusses the related work. We conclude the paper in Section~\ref{sec:conclusion}.

\section{Background and Conceptual Framing}
\label{sec:background}

\subsection{Technical Debt and Maintenance Prioritization}

Software maintenance and evolution require continuous prioritization~\cite{de2008empirical}. Since teams cannot address all quality issues at once, they must focus limited effort where it yields the greatest practical benefit~\cite{kim2012field}. Although \emph{technical debt} captures how suboptimal %technical
decisions increase future maintenance costs, it remains a broad concept. In practice, teams need support not only to identify technical weaknesses, but also to determine which ones most strongly hinder ongoing development~\cite{kim2012field}. Software analytics provide useful signals, including code quality indicators, hotspots, and ownership-related views~\cite{nagappan2005use, zimmermann2005mining, bird2011don}. These can reveal technically problematic, change-prone, or organizationally risky parts of a system~\cite{nagappan2005use, cataldo2008socio}. However, such signals often remain fragmented, leaving teams without an integrated view of which parts impose the greatest practical burden on maintenance and evolution.

\subsection{CodeScene CodeHealth and Hotspots}

%ARDA
%%%%%CodeHealth™ (CH) is a quality metric used in the CodeScene software engineering intelligence platform. Its goal is to capture how cognitively difficult it is for human developers to comprehend source code. CodeScene identifies \textit{code smells} and combines the number and severity of detected smells into a file-level score from 1 to 10. Lower scores indicate higher cognitive load for humans, i.e., a higher maintenance effort. CodeScene presents LoC-weighted averages of CH at the project-level, as well as averages for architectural components and user-selected sets of files. We have previously validated CH through its association with file-level defect density and development time~\cite{tornhillCodeRedBusiness2022,borgIncreasingNotDiminishing2024}, as well as through a comparative study on a public benchmark~\cite{borg2024ghost}. CodeScene uses the construct \textit{hotspots} to identify code with the highest development activity. In active software projects, the number of revisions per file typically follows a heavily skewed distribution, where a small subset of files accounts for a large proportion of changes. This subset is determined using a knee point analysis of the change frequency distribution, separating frequently modified files from the long tail of rarely changed files. The average CH of the files flagged as hotspots is referred to as the Hotspot~CH (HS\_CH).

CodeHealth™ (CH) is a quality metric in the CodeScene SE intelligence platform~\cite{codescene} intended to capture how cognitively difficult source code is for developers to understand. It combines the number and severity of detected \textit{code smells} into a file-level score from 1 to 10, where lower scores indicate higher cognitive load and maintenance effort. CodeScene also reports LoC-weighted CH averages at the project level, for architectural components, and for user-selected file sets. Prior work has validated CH through its association with file-level defect density and development time~\cite{tornhillCodeRedBusiness2022,borgIncreasingNotDiminishing2024}, as well as in a comparative study on a public benchmark~\cite{borg2024ghost}. CodeScene uses \emph{hotspots} to identify the code with the highest development activity. In active projects, file revisions typically follow a highly skewed distribution, where a small subset of files accounts for a large share of changes. CodeScene identifies this subset through knee-point analysis of the change-frequency distribution, separating frequently modified files from the long tail of rarely changed ones. The average CH of the flagged hotspot files is referred to as the Hotspot CH (HS\_CH).

\subsection{Technical Debt Friction}
\label{sec:friction}

\emph{Technical debt friction} denotes the practical resistance parts of a system impose on ongoing maintenance and change~\cite{CodeSceneTechnicalDebtFriction}. This resistance may stem from structural weaknesses, frequent changes, unstable dependencies, refactoring difficulty, or socio-technical factors such as fragmented ownership and knowledge bottlenecks~\cite{zimmermann2005mining, kim2012field, bird2011don, cataldo2008socio,10.1145/3691620.3695508}. The concept shifts attention from isolated quality deficiencies to their concrete impact on day-to-day software work. Since maintenance difficulties typically arise from the interplay of technical and organizational conditions rather than a single cause~\cite{bird2011don, cataldo2008socio}, technical debt friction aims to capture this broader burden and support more effective maintenance prioritization~\cite{CodeSceneTechnicalDebtFriction}.

We calculate friction at two levels. 
At the project level, friction is used for characterizing and comparing projects, resulting in a score on a bounded 0–99 scale as follows:

\[
\text{Fric}_{\text{proj}} = (10 - \text{HS\_CH}) + \left(100 - 10 \cdot \frac{\sum_{f \in F} CH_f \cdot Rev_f}{\sum_{f \in F} Rev_f} \right)
\]

where \( f \in F \) denotes the individual files in the project, \( CH_f \) is the CodeHealth of file \( f \), and \( Rev_f \) is the number of revisions to file \( f \). We refer to the first term as the \textit{start friction} and the second term as the \textit{dynamic friction}. Note that the activity-weighted term includes only files with at least one revision during the analysis period (default value: 1~year). If no file has been revised, project-level friction is undefined. We discuss project-level friction scores in Section~\ref{sec:res_rq3}.

At the file-level, friction is calculated for localization and visualization in the interactive user interface. This score also combines relative revision activity and CH, while reserving part of the visual scale to emphasize leading hotspots. In Section~\ref{sec:res_rq2} we discuss file-level friction scores.

\subsection{Relation to Adjacent Concepts}

Technical debt friction is related to, but distinct from, several established concepts. It goes beyond \emph{code health} by emphasizing the practical burden of evolving code~\cite{CodeSceneTechnicalDebtFriction}. It also differs from \emph{hotspots}, which in CodeScene identify files with high development activity regardless of code quality. Earlier platform versions combined hotspots with low-code health to flag a small, actionable subset of files, whereas technical debt friction provides a more fine-grained signal of how development activity interacts with maintainability concerns across the codebase. Likewise, \emph{change coupling} may reveal one source of maintenance difficulty, but does not itself capture the broader resistance experienced in practice~\cite{zimmermann2005mining}. Nor is high friction equivalent to a refactoring decision, which also depends on strategic importance, business constraints, and implementation cost~\cite{kim2012field}. Finally, technical debt friction also has a socio-technical dimension: maintenance burden depends not only on code structure, but also on how knowledge, ownership, and coordination are distributed across the team~\cite{bird2011don, cataldo2008socio}. We therefore view technical debt friction as an integrative concept for reasoning about maintenance burden through the combined lens of change activity and maintainability.

\subsection{Study Focus}

%%%%%The goal of this paper is not to argue that technical debt friction replaces existing notions such as technical debt, code health, or hotspots. Rather, we position it as a practically oriented concept whose value depends on whether practitioners find it understandable, credible, and useful in real software development contexts. Accordingly, our study investigates whether practitioners interpret friction in meaningful ways, whether friction-related analyses align with their perceived maintenance pain points, and whether the concept supports actionable reasoning about maintenance prioritization. In this sense, technical debt friction is treated as a decision-support lens for identifying where a system most strongly resists change and where improvement efforts may have the highest practical payoff.

%ARDA
%%%%%The goal of this paper is not to present technical debt friction as a replacement for existing notions such as technical debt, code health, or hotspots. Instead, we treat it as a practice-oriented concept whose value depends on whether practitioners find it understandable, credible, and useful in real development contexts. Accordingly, our study examines whether practitioners interpret friction meaningfully, whether friction-related analyses align with their perceived development pain points, and whether the concept supports actionable maintenance prioritization. In this sense, technical debt friction is treated as a decision-support lens for identifying where a system most strongly resists change and where improvement efforts may yield the greatest practical benefit.

Our goal is not to position technical debt friction as a replacement for concepts such as technical debt, code health, or hotspots. Instead, we treat it as a practice-oriented concept whose value depends on whether practitioners find it understandable, credible, and useful in real development contexts. Accordingly, we examine whether practitioners interpret friction meaningfully, whether friction-related analyses align with perceived development pain points, and whether the concept supports actionable maintenance prioritization. In this sense, debt friction serves as a decision-support lens for identifying where a system most strongly resists change and where improvement efforts may yield the greatest practical benefit.

\section{Research Design}
\label{sec:research-design}
We conducted a multi-case exploratory case study~\cite{yin1989case} with practitioner-based concept validation. %to investigate how practitioners interpret technical debt friction, how it aligns with perceived maintenance pain points, and whether it supports actionable maintenance prioritization. 
While RQ1 and RQ2 are grounded in the practitioner interviews and walkthrough sessions, RQ3 complements this design with an exploratory project-level analysis of friction distributions.

\subsection{Analyzed Projects}
\label{sec:analyzed-projects}

%ARDA
%%%%%This study involved CodeScene-based analyses of three software products from industry, with Products A and C each represented by two versions (Table~\ref{tab:Products}). The cases span modern IoT and data-integration platforms as well as a large legacy monolith, and vary considerably in size (117K--2.8M LOC) and architectural complexity (19--76 components). All product versions were analyzed with CodeScene, with three to four interviewees per case. The largest, \textbf{Product C}, is a monolithic system with 2.8 million (M) LOC, while \textbf{Products A and B} have more modern code bases. Product A is a heavy UI-driven application that connects to a backend IoT system. Product B is a modern .Net project, with some mixes of JavaScript and Python. CodeScene~\cite{codescene} assessed key code quality metrics, revealing that, despite their smaller size, \textbf{A and B} have more architectural components than C, indicating greater modularity. All three products have minor integrations of \textbf{Python}.

This study analyzed three industrial software products with CodeScene, with Products A and C each represented by two versions (Table~\ref{tab:Products}). The cases span modern IoT and data-integration platforms as well as a large legacy monolith, and vary substantially in size (117K--2.8M LOC) and architectural complexity (19--76 components). %All product versions were analyzed with CodeScene~\cite{codescene}, with three to four interviewees per case. 
Product C is the largest, a 2.8M-LOC monolithic system, whereas Products A and B are more modern codebases. Product A is a UI-intensive application connected to a backend IoT system, and Product B is a modern .NET system with some JavaScript and Python. %CodeScene showed that, 
Despite their smaller size, Products A and B contain more architectural components than Product C, indicating greater modularity. The products include minor Python integrations.

\begin{table*}[h!]
\scriptsize
\caption{Summary of industrial projects used in the study.}
\centering
\label{tab:Products}
\vspace*{-0.8em}
\begin{tabular}{ l  m{5.3cm}  m{3.5cm}  m{2.05cm}  m{0.5cm}  m{1.3cm}  m{1.55cm} }
%\hline
\toprule
\textbf{Product} & \textbf{Short description} & \textbf{Programming languages} & \textbf{\#Arch Components} & \textbf{LOC} & \textbf{Tool(s) used} & \textbf{\# Interviewees} \\
%\hline
\toprule
A v1 & A modern software platform for an IoT system & TypeScript, JavaScript, Python & 21 & 210K & CodeScene & 4 \\
\rowcolor{black!6}
A v2 & A modern software platform for an IoT system & TypeScript, JavaScript, Python & 21 & 235K & CodeScene & 4 \\
%\hline
B & A modern software supporting data integration & C\# .Net 8, JavaScript, Python & 76 & 117K & CodeScene & 3 \\
\rowcolor{black!6}
%\hline
C v1 & A big monolith with a lot of legacy code  & C\#, C++, Python & 19 & 2.8M & CodeScene & 3 \\
C v2 & A big monolith with a lot of legacy code & C\#, C++, Python & 19 & 910k & CodeScene & 3 \\
%\hline
% \rowcolor{black!6}
% D & A cloud-based solution & JavaScript, C\#, TypeScript & 71 & 2.3M & CodeScene & ? \\
%\hline
\toprule
\end{tabular}
\vspace*{-1.9em}
\end{table*}

\subsection{Data Collection - Tools}
\label{sec:data-collection-tools}

%ARDA
%%%%%We used \emph{CodeScene}~\cite{codescene} as the main analysis tool because it provides an integrated set of code, evolutionary, and socio-technical views relevant to our study. In particular, it offers \emph{Code Health}, \emph{Hotspots}, and \emph{Technical Debt Friction}, enabling the inspection of both structural code quality and the concentration of development activity across the system. A key reason for choosing CodeScene is that it directly provides \emph{Technical Debt Friction} as an analysis view. This was important because our goal was not to construct a new friction metric, but to investigate whether this tool-supported notion is meaningful and useful in industrial practice. We therefore used CodeScene as an analysis and discussion aid, using its visualizations and prioritized results to ground practitioner discussions about maintenance pain points, refactoring needs, and the practical burden of change.

We used \emph{CodeScene}~\cite{codescene} as the main analysis tool because it provides an integrated set of code, evolutionary, and socio-technical views relevant to our study, including \emph{Code Health}, \emph{Hotspots}, and \emph{Technical Debt Friction}. A key reason for this choice is that CodeScene directly provides \emph{Technical Debt Friction} as an analysis view. This was important because our goal was not to construct a new friction metric, but to investigate whether this tool-supported notion is meaningful and useful in industrial practice. We therefore used CodeScene as an analysis and discussion aid to ground practitioner discussions about maintenance pain points, refactoring needs, and the practical burden of change.

\begin{figure}[t]
    \centering
    \includegraphics[width=\linewidth]{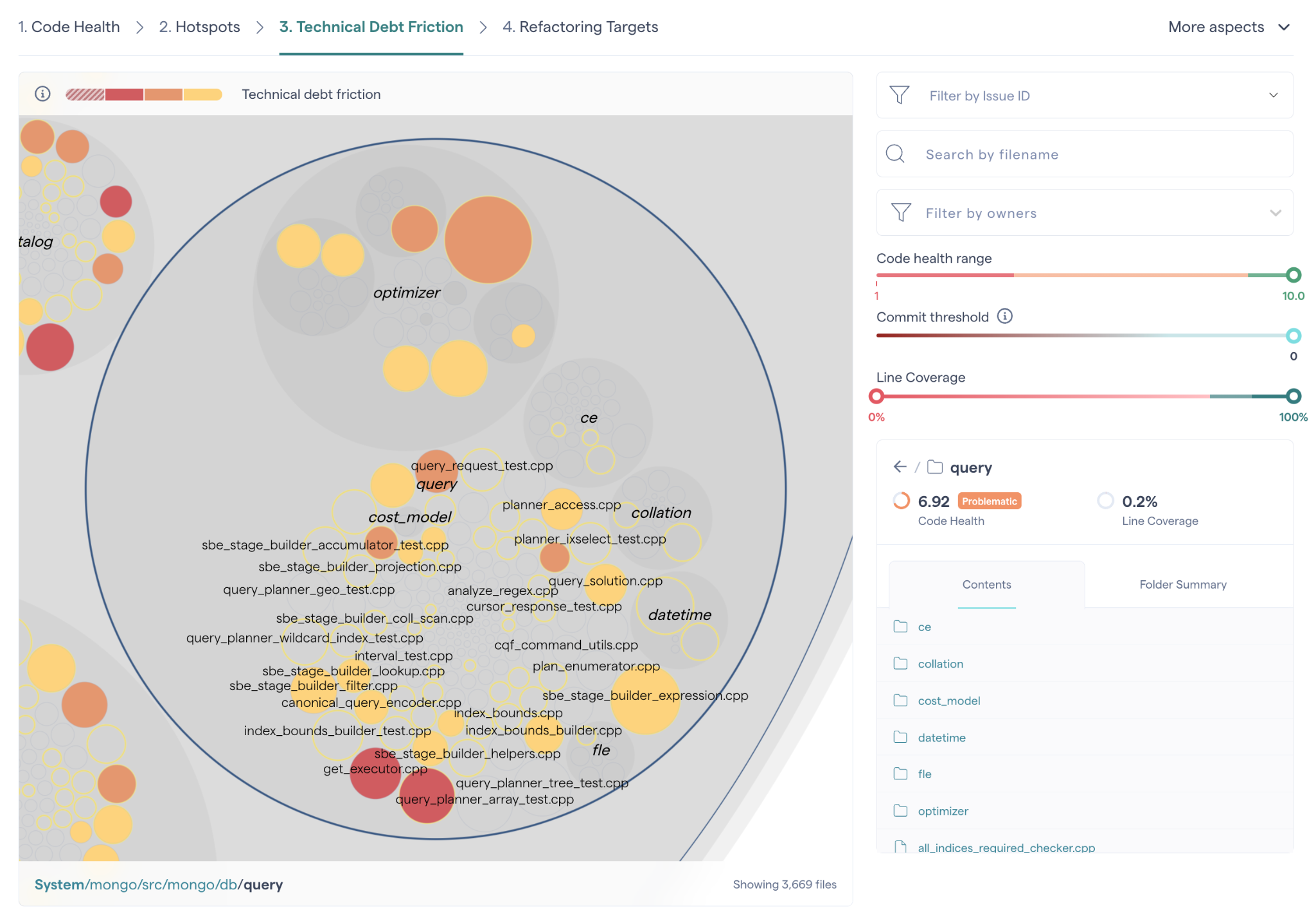}

    \vspace*{-0.8em}

    \caption{Example technical debt friction view in CodeScene. It highlights parts of the codebase where low code health overlaps with development activity.} %%%, helping identify areas likely to impose a high maintenance burden. %Larger circles indicate higher relative significance, while warmer colors denote more problematic code health.}
    \label{fig:td_friction}

    \vspace*{-1.4em}

\end{figure}

%ARDA
%%%%Figure~\ref{fig:td_friction} illustrates an example \emph{Technical Debt Friction} view in CodeScene. The visualization highlights parts of the codebase where low code health coincides with development activity, thereby helping identify areas likely to impose a high practical burden on maintenance and evolution. Larger circles represent files or components with greater relative significance in the analysis, while color encodes code health, with warmer colors indicating more problematic areas. In this way, the view supports identifying locations where technical debt is most likely to affect day-to-day development work and where improvement efforts may have the greatest practical payoff.

%ARDA
%%%%%Figure~\ref{fig:td_friction}  illustrates an example \emph{Technical Debt Friction} view in CodeScene. The visualization highlights parts of the codebase where current code health interacts with recent development activity, thereby helping identify areas likely to impose a high practical burden on maintenance and evolution. Circle size reflects the LOC of individual files, while larger gray directory circles indicate directories containing more files. Color encodes \emph{Technical Debt Friction}, which combines present code health with commit activity during the last 12 months, with warmer colors indicating higher friction. In this way, the view helps identify locations where technical debt is most likely to affect day-to-day development work and where improvement efforts may have the greatest practical payoff.

Figure~\ref{fig:td_friction} shows an example \emph{Technical Debt Friction} view in CodeScene. It highlights parts of the codebase where current code health and recent development activity combine to create likely maintenance burden. Circle size reflects the LOC of individual files, while larger gray directory circles indicate directories with more files. Color encodes \emph{Technical Debt Friction}, combining present code health with commit activity over the last 12 months, with warmer colors indicating higher friction. The view thus helps identify locations where technical debt is most likely to affect day-to-day development work and improvement efforts may have the greatest practical payoff.

\subsection{Data Collection - Survey with Practitioners}
\label{sec:data-collection-survey}

%ARDA
%%%%%%To analyze each software product, we installed CodeScene (being dockerized) on Linux virtual machines (VMs) provided by the companies. Product owners cloned their codebases onto VMs with restricted researcher access for tool execution. We ran the tools, documented key findings, and held online seminars with developers, architects, and managers. CodeScene successfully worked across all products (Table~\ref{tab:Products}). %Product A is a modern software platform for an IoT system. Product B is a modern software product supporting data integration. Product C is a large monolith with a substantial amount of legacy code, currently largely in maintenance mode, with some parts no longer actively used.
%%%Product A: TODO
%%%Product B: TODO. 
%%%Product C is a big monolith with a lot of legacy code mainly in maintenance mode, some parts are not really used anymore. 

%ARDA
%%%%%The interviews were conducted as part of a case study exploring how practitioners perceive and use socio-technical analysis results in their daily work. We used a semi-structured interview protocol built around CodeScene, a software analytics tool. During each interview, the researcher shared their screen and navigated the CodeScene reports for the participant’s project while prompting discussion.

To analyze each software product, we deployed the dockerized CodeScene tool on company-provided Linux virtual machines (VMs). Product owners cloned their codebases onto these VMs, and researchers were given restricted access to run the tool. We documented the results and conducted online seminars with developers, architects, and managers; CodeScene worked successfully across all products (Table~\ref{tab:Products}). The interviews were conducted as part of a case study on how practitioners perceive and use socio-technical analysis results in daily work. Using a semi-structured protocol, the researcher shared their screen, navigated the CodeScene reports for the participant’s project, and prompted discussion.

%%%%%\subsubsection{Interview Steps}

%ARDA
%%%%%Across all interviews, we followed the same high-level sequence: (i) introduction and overview of CodeScene results, (ii) exploration of code health, (iii) hotspot analysis, (iv) friction and change coupling, and (v) evaluation of friction and overall usefulness.

Across all interviews, we followed the same sequence: (i) introduction and overview of CodeScene results, (ii) code health, (iii) hotspots, (iv) friction and change coupling, and (v) evaluation of friction and overall usefulness.

%%%\begin{itemize}
%%%    \item Introduction and overview of CodeScene results
%%%    \item Exploration of code health
%%%    \item Hotspot analysis
%%%    \item Friction and change coupling
%%%    \item Evaluation of friction and overall usefulness
%%%\end{itemize}

\begin{table}[t]
\scriptsize
\centering
\caption{Number of participants in our study.}
\label{tab:participants}
\vspace*{-1.2em}
\begin{tabular}{p{0.35\linewidth} c c c }
\toprule
\textit{\textbf{Team}} & \textit{\textbf{Developer}} & \textbf{\textit{Architect}}  &\begin{tabular}[c]{@{}l@{}}\textit{\textbf{Manager}}\end{tabular}\\ \toprule
\textit{Product Team A} & 2 & 1  & 1\\ %\hline 
\rowcolor{black!6}
\textit{Product Team B} & 1 & 1  & 1\\ %\hline 
\textit{Product Team C} & 1 & 1  & 1\\ 
%\hline 
% \rowcolor{black!6}
% \textit{CompanyY.TeamD} & - & ?  & ?\\ 
% \bottomrule
%\rowcolor{black!6}
 \textbf{Total Individual Count} & 4 & 3 & 3\\ \midrule
\end{tabular}
\vspace*{-2.0em}
\end{table}

\subsubsection{Introduction and Overview of CodeScene Results}
%ARDA
%%%%Each interview began with a short explanation of the study purpose, then the interviewer opened the project’s CodeScene dashboard and briefly introduced the main views (overall status, code health, hotspots, and friction). 

Each interview began with a short explanation of the study purpose. The interviewer then opened the project's CodeScene dashboard and briefly introduced the main views, including overall status, code health, hotspots, and friction.

\subsubsection{Exploration of Code Health}

%ARDA
%%%%%The interviewer then focused on the \emph{Code Health} view. A file highlighted in orange-red (low code health) was selected directly from the CodeScene interface. For this file, the interviewer presented CodeScene’s warnings and violations (e.g., many conditionals, complex methods, cyclomatic complexity and its evolution over time) and displayed the corresponding source code when requested. Participants were asked to comment on these indicators, to what extent they agreed with them, and how they related to their own assessment of the code.

The interviewer focused on the \emph{Code Health} view and selected a file with low code health in the interface. CodeScene's warnings and violations (e.g., many conditionals, complex methods, cyclomatic complexity and its evolution) were discussed, and the corresponding source code was shown when needed. Participants were asked whether these indicators matched their assessment of the code.

\subsubsection{Hotspot Analysis}
%ARDA
%%%%Next, the \emph{Hotspots} view was used, showing the activity (commits) in the last 12 months. The interviewer highlighted files that combined high change frequency and invited participants to reflect on whether these hotspots matched their expectations and experiences of frequently touched parts of the system.

The interviewer then used the \emph{Hotspots} view, showing commits over the last 12 months, and highlighted files with high change frequency. Participants reflected on whether these hotspots matched their expectations and experience of frequently modified parts of the system.

\subsubsection{Friction}

%ARDA
%%%%%The interviewer introduced the notion of friction as high development activity in low-quality code. Using CodeScene’s friction view, one file with high friction was selected. Participants were asked whether they recognized this file as a source of friction, whether it had caused issues in practice, and whether they had refactored or would consider refactoring it. 

The interviewer introduced friction as high development activity in low-quality code. Using CodeScene's friction view, one highly flagged file was selected (multiple flagged files were selected for each product), and participants were asked whether they recognized it as a source of friction, whether it had caused problems in practice, and whether they had refactored or would consider refactoring it.

\subsubsection{Friction and Change Coupling}

%ARDA
%%%%%The interviewer also displayed dependencies and change coupling information for this file, asking participants how well these patterns reflected the actual architecture and coordination needs and whether such information would be useful for planning refactorings.

The interviewer also showed dependencies and change-coupling information for the same file, asking how well these patterns reflected the actual architecture and coordination needs and whether such information would be useful for planning refactorings.

\subsubsection{Evaluation of Friction and Overall Usefulness}
%ARDA
%%%%%In the final part of the interview, participants were asked to assess the usefulness of the friction view and the CodeScene reports. They were invited to comment on which views and metrics they found most or least useful, how they might use such analyses to prioritize refactoring and maintenance, and whether they saw value in using these visualizations to communicate technical debt and quality concerns to other stakeholders.

In the final part, participants assessed the usefulness of the friction view and the CodeScene reports overall. They commented on which views and metrics they found most or least useful, how they might use such analyses to prioritize refactoring and maintenance, and whether these visualizations could help communicate technical debt and quality concerns to others. %stakeholders.

\subsubsection{Data Collection and Analysis}
%ARDA
%%%%%Interviews were conducted remotely via video conferencing, with audio and screen sharing recorded subject to consent. The interviewer took notes during each session, and recordings were transcribed verbatim. We applied thematic coding~\cite{cruzes2011recommended}, to the transcripts, focusing on (i) how practitioners interpret automated quality indicators, (ii) how they relate hotspots and friction to their lived experience of the codebase, and (iii) under what conditions they consider such metrics actionable. The shared protocol and sequence of tool views ensured comparability across interviews while still allowing for open-ended, in-depth discussion. Table~\ref{tab:participants} presents the study participants. %%%%%from the product teams. The last row of Table~\ref{tab:participants} shows the total unique participants in the study.

Interviews were conducted remotely via video conferencing, with audio and screen sharing recorded subject to consent. The interviewer took notes in each session, and recordings were transcribed verbatim. We applied thematic coding~\cite{cruzes2011recommended}, focusing on (i) how practitioners interpret automated quality indicators, (ii) how they relate hotspots and friction to their lived experience of the codebase, and (iii) under what conditions they consider such metrics actionable. The shared protocol and tool sequence ensured comparability across interviews while allowing open-ended, in-depth discussion. Table~\ref{tab:participants} presents the study participants.

\section{Results}
\label{sec:results}

%%%%This section discusses the results of our case studies, addressing, in turn, each of the RQs. 

\subsection{RQ1: Practitioner Interpretation of Technical Debt Friction}

%ARDA
%%%%To address \textit{RQ1}, we analyzed how practitioners discussed and assessed \emph{Technical Debt Friction} during the study sessions across the three products. Our analysis focused on whether the concept was perceived as meaningful in practice, what kinds of files and maintenance situations practitioners associated with friction, and which contextual factors shaped their interpretation of the metric. %%%%Overall, we found that practitioners generally treated friction as a useful indicator of maintenance burden, but interpreted it through their existing knowledge of the codebase, the role of the file, and the broader development context rather than as a standalone signal.

To address \textit{RQ1}, we analyzed how practitioners discussed and assessed \emph{Technical Debt Friction} across the three products, focusing on whether they found it meaningful in practice, which files and maintenance situations they associated with it, and which contextual factors shaped their interpretation.

%ARDA
%%%%Across the three products, practitioners generally found \emph{Technical Debt Friction} useful when the highlighted files corresponded to maintenance-relevant parts of the codebase. At the same time, the sessions showed that friction was not interpreted as a standalone or self-sufficient signal. Instead, practitioners read it through knowledge of the codebase, the role of the file, the surrounding architecture, and the current development context. We identified four main findings.

Across the products, participants generally found \emph{Technical Debt Friction} useful when the flagged files corresponded to maintenance-relevant parts of the codebase. At the same time, they did not treat friction as a standalone signal, but interpreted it through knowledge of the codebase, file role, architecture, and current development context. We identified four findings.

\paragraph{Friction often aligned with existing practitioner knowledge}
%ARDA
%%%%A first finding is that high-friction files frequently overlapped with files that practitioners already considered problematic. In Product C, one flagged file was recognized by developers as a likely source of friction because of its deep nesting, brain class characteristics, and low cohesion. In Product A, files identified as having very high friction were also associated with bumpiness and nested complexity, and one of these files was later refactored, reportedly reducing friction and improving code health. These examples indicate that friction often highlighted code practitioners already associated with maintenance difficulty. This overlap is important because it suggests that the metric resonated with lived experience rather than surfacing only implausible or irrelevant candidates.

High-friction files frequently overlapped with files that participants already considered problematic. In Product C, one flagged file was recognized as a likely source of friction because of deep nesting, brain class characteristics, and low cohesion. In Product A, highly flagged files were similarly associated with bumpiness and nested complexity, and one was later refactored, reducing friction and improving code health. This overlap suggests that friction often resonated with lived experience rather than surfacing only implausible candidates.

\paragraph{Friction could also surface less expected candidates, but these required contextual interpretation}
%ARDA
%%%%%A second finding is that the sessions also surfaced files that were not initially expected by practitioners, although these cases still required contextual interpretation. In Product C, another flagged file showed similar issues related to low cohesion, bumpiness, and nested complexity, and developers confirmed that it was functionally overloaded, with too much diverse functionality concentrated in a single file. This file was therefore seen as a plausible refactoring candidate even though it had not initially been highlighted by the developers themselves. At the same time, not all unexpected files were considered equally important. In Product B, one file with lower friction than the outliers was still recognized as a complex file with relatively frequent changes, but the developers considered the observed issues less severe. These results suggest that practitioners did not assess friction mechanically; instead, they used it as a prompt for further inspection and validation.

The sessions also highlighted files that practitioners had not initially expected, though these still required validation. In Product C, one flagged file showed low cohesion, bumpiness, and nested complexity, and participants confirmed that it was functionally overloaded, making it a plausible refactoring candidate despite not having been highlighted beforehand. At the same time, not all unexpected files were equally important. In Product B, where the standout files showed lower overall friction than in the other products, one flagged file was still recognized as complex and frequently changed, but the participants considered its issues less severe in context due to its role. These cases show that participants did not interpret friction mechanically, but used it as a prompt for further inspection and validation.

\paragraph{Practitioners did not treat friction as an automatic refactoring signal}
%ARDA
%%%%%A third finding is that practitioners interpreted friction as an indicator of burden, but not necessarily as a direct trigger for action. In several discussions, high friction was read as a negative condition, yet not every high-friction file was seen as a high-priority refactoring target. In Product C, one high-friction file was explicitly treated as lower priority because it was relatively isolated and had limited coupling to the rest of the codebase. Developers noted that a similar score in a more central component would likely have led to a different assessment. Likewise, in Product B, some highly flagged files were test files. Although the underlying issues were acknowledged, developers considered those files less relevant for their particular use case and therefore lower priority than production files with similar scores. These examples show that practitioners interpreted friction as a useful warning signal, but one that still required prioritization through contextual judgment.

They interpreted friction as an indicator of burden, but not necessarily as a direct trigger for action. In Product C, one high-friction file was treated as lower priority because it was isolated and had no change coupling to the rest of the codebase; participants noted that the same score in a more central component would likely have been judged differently. Likewise, in Product B, some highly flagged files were test files. Although the underlying issues were acknowledged, participants considered them less relevant %to their use case 
and therefore lower priority than production files with similar scores. These examples show that friction was treated as a useful warning signal, but one that still required contextual prioritization.

\paragraph{Product context strongly shaped the meaning of friction}
%%%%A fourth finding is that practitioners consistently interpreted friction through the broader product and development context. Two contextual factors were especially salient. First, developers pointed out that friction is partly driven by recent change frequency and therefore may miss files that are rarely modified but known or suspected to be difficult and time-consuming when change is required. Second, they noted that active refactoring can itself inflate friction, since files undergoing cleanup often still exhibit poor code health while also receiving many commits. This was particularly visible in Product C and suggests that friction must be interpreted together with knowledge of ongoing work. More generally, participants repeatedly emphasized that the practical relevance of friction depends on the role of the file in the product, whether it belongs to production or test code, how strongly it is coupled to other parts of the codebase, and if recent activity reflects normal maintenance or temporary restructuring.

Participants consistently interpreted friction through the broader product and development context. Two factors were especially salient. First, because friction is partly driven by recent change frequency, it may miss files that are rarely modified but still difficult and time-consuming to change. Second, active refactoring can itself inflate friction, since files under cleanup may still have poor code health while receiving many commits. This was particularly visible in Product C and suggests that friction must be interpreted together with knowledge of ongoing work. More generally, participants emphasized that the practical relevance of friction depends on file role, whether it belongs to production or test code, its coupling to other parts of the codebase, and whether recent activity reflects routine maintenance or temporary restructuring.

\begin{table*}[t]
\centering
\caption{Summary of files examined for RQ2 and their observed alignment with maintenance relevance.}
\vspace*{-1.0em}
\label{tab:rq2_alignment_summary}
\scriptsize
\begin{tabular}{p{0.55cm} p{1.85cm} p{2.45cm} p{4.60cm} p{3.6cm} p{1.7cm} p{0.85cm}}
\toprule
\textbf{Product} & \textbf{Flagged file} & \textbf{Why highlighted} & \textbf{Practitioner assessment} & \textbf{Observed later outcome} & \textbf{Alignment} & \textbf{Friction} \\
\midrule

A v1 
& File 1 
& Low code health, high friction; structural issues 
& Matched known problem areas; considered maintenance-relevant 
& Later refactored; friction decreased and code health improved 
& Strong 
& 43\% \\

\rowcolor{black!6}
A v1 
& File 2 
& High commits/year, high friction, low code health 
& Recognized as problematic and architecturally important 
& Refactoring deferred due to importance, but still considered relevant 
& Strong 
& 22\% \\

A v1 
& File 3 / 4 / 5 
& High friction with varying structural quality 
& Seen as relevant, but importance differed by file role and context 
& Some remained relevant; others reflected more temporary activity 
& Partial 
& 16–19\% / 6\% \\

\rowcolor{black!6}
A v2 
& Previously flagged File 1 
& Reflected from earlier snapshot 
& Still considered relevant 
& Refactored by follow-up; friction reduced and code health improved 
& Strong 
& 4\% \\

A v2 
& Previously flagged File 2 / 3 / 4 
& Reflected from earlier snapshot 
& Relevance depended on whether changes reflected restructuring or persistent design issues 
& Some improved; some remained under active development 
& Partial 
& 8\% / 4\% / 4\% \\

\rowcolor{black!6}
B 
& File 1 
& Complex logic, frequent changes, high friction 
& Recognized as relevant, though participants argue for necessary friction in this file %issues seen as less severe than metric suggested 
& Remained a plausible improvement candidate 
& Moderate 
& 23\% \\

B 
& File 2 
& High friction and quality concerns 
& Issues acknowledged, but lower priority because it was test code 
& Would be refactored only if needed in practice 
& Weak / context-dependent 
& 17\% \\

\rowcolor{black!6}
B 
& File 3 / 4 / 5 
& Flagged due to friction 
& Considered less relevant than production code; some likely influenced by recent activity 
& Limited practical importance for prioritization 
& Weak 
& 19\% /6\% \\

C v1 
& File 1 
& Very high friction, deep nesting, brain class, low cohesion 
& Strongly matched practitioner expectations; considered difficult and burdensome 
& Reasonable refactoring candidate 
& Strong 
& 45\% \\

\rowcolor{black!6}
C v1 
& File 2 
& Similar issues, but more isolated in architecture 
& Acknowledged as problematic, but lower priority due to weaker coupling and local impact 
& No immediate action planned 
& Partial 
& 33\% \\

C v1 
& File 3 
& High friction during active rework 
& Developers believed refactoring was ongoing 
& Friction decreased, but remained influenced by high change activity 
& Partial 
& 33\% \\

\rowcolor{black!6}
C v2 
& Previously flagged files from v1 
& Revisited in follow-up snapshot 
& Some remained relevant, but interpretation depended on whether activity reflected ongoing maintenance or completed rework 
& Mixed: some improved, others stayed similar 
& Partial 
& 38\%/ 9\%/ 14\% \\

\bottomrule
\end{tabular}
\vspace*{-1.9em}
\end{table*}
 
\begin{tcolorbox}[boxsep=2pt,left=2pt,right=2pt,top=2pt,bottom=2pt]

\textbf{RQ1 Conclusion.} Practitioners interpreted \emph{Technical Debt Friction} as a meaningful but context-dependent indicator of maintenance burden. High-friction files were often aligned with existing practitioner knowledge of problematic code, and in some cases the metric also surfaced less expected but plausible candidates for further inspection. However, practitioners did not treat friction as a standalone refactoring signal. Instead, they interpreted it through architectural centrality, file type, ongoing development activity, and their broader understanding of the product. These findings indicate that friction is most useful when combined with practitioner knowledge rather than used in isolation.

\end{tcolorbox}

\subsection{RQ2: Alignment with Maintenance-Relevant Areas}
\label{sec:res_rq2}

%%%To address RQ2, we examined whether files highlighted by \emph{Technical Debt Friction} corresponded to areas that later proved maintenance-relevant in practice. We did this by comparing the initial CodeScene snapshot with subsequent changes in code health and friction, developers' assessments during the interviews, and later refactoring or maintenance activity. Overall, we found that friction aligned well with several files that later received attention or were already recognized as problematic, but that this alignment was strongly shaped by file role, architectural centrality, and ongoing development context.

%ARDA
%%%%%To address \textit{RQ2}, we examined whether files highlighted by \emph{Technical Debt Friction} corresponded to areas that later proved maintenance-relevant in practice. We did this by comparing the initial CodeScene snapshot with subsequent changes in code health and friction, developers' assessments during the interviews, and later refactoring or maintenance activity. Table~\ref{tab:rq2_alignment_summary} summarizes the main files examined in this analysis, their practitioner assessment, and their observed later alignment with maintenance relevance. Overall, we found that friction aligned well with several files that later received attention or were already recognized as problematic, but that this alignment was strongly shaped by file role, architectural centrality, and ongoing development context.

To address \textit{RQ2}, we examined whether files highlighted by \emph{Technical Debt Friction} corresponded to areas that later proved maintenance-relevant in practice. We compared the initial CodeScene snapshot with subsequent changes in code health and friction, participants’ assessments during the interviews, and later refactoring or maintenance activity. Table~\ref{tab:rq2_alignment_summary} summarizes the main files examined, their practitioner assessment, and their observed later alignment with maintenance relevance. It reports representative friction values for selected files to provide a quantitative indication of signal strength while focusing on alignment rather than thresholds. To classify alignment between \emph{Technical Debt Friction} and maintenance relevance, we used a qualitative, evidence-based scheme. Alignment was \emph{strong} when practitioner recognition, technical indicators, and observed maintenance activity converged; \emph{moderate} when recognition and indicators aligned but improvement evidence was limited; \emph{partial} when only some evidence supported relevance; and \emph{weak} or \emph{context-dependent} when flagged files were judged lower priority (e.g., test code, peripheral components, or temporary spikes). %%%%These categories capture the degree of convergence between tool signals, practitioner interpretation, and observed activity, rather than defining strict thresholds.

\paragraph{Friction often aligned with files that later received maintenance attention}
%ARDA
%%%%%A first finding is that several files highlighted by friction were later refactored, improved, or remained under active maintenance attention. In Product A-v1, one highly flagged file was later refactored, which reduced friction and improved code health. A second highly flagged file in the same product was also recognized as problematic, although future refactoring was deferred because of its importance. In Product A-v2, one of the files identified in the earlier snapshot had been successfully refactored by the time of the follow-up, leading to improved code health and lower friction. In Product C-v1, one flagged file had already been refactored during the observation period, which reduced friction even though change activity remained high. These examples indicate that friction frequently identified files that either became targets for improvement or were already part of ongoing maintenance work.

Several files highlighted by friction were later refactored, improved, or remained under active maintenance. In Product A-v1, one highly flagged file (43\% friction) was later refactored, reducing friction and improving code health, while another file (22\% friction) was also recognized as problematic but deferred because of its low importance. In Product A-v2, one previously flagged file had been successfully refactored by the follow-up, again leading to improved code health and lower friction (e.g., from 22\% to 8\%). In Product C-v1, one flagged file (33\% friction) had already been refactored during the observation period, reducing friction even though change activity remained high. These examples indicate that friction frequently identifies files that later become targets for improvement or remain central to ongoing maintenance work.

\paragraph{Alignment was strongest for central production code with persistent structural problems}
%ARDA
%%%%A second finding is that alignment was clearest when flagged files combined structural issues with architectural or functional importance. In Product A-v1, files with low code health were considered meaningful because they were central to the product and exhibited issues such as bumpiness, nested complexity, and low cohesion. In Product B, a highly flagged production file with complex logic and frequent changes was seen as important, even if developers considered some of its issues less severe than the metric suggested. Similarly, in Product C-v1, one flagged file was explicitly recognized as difficult because of its large size and nested conditionals, and developers considered it a reasonable candidate for refactoring. These cases suggest that friction aligned most clearly with maintenance-relevant areas when the file was both structurally problematic and important to everyday development work.

Alignment was clearest when flagged files combined structural issues with architectural or functional importance. In Product A-v1, files with low code health and moderate-to-high friction (e.g., 16–19\%) were considered meaningful because they were central to the product and showed bumpiness, nested complexity, and low cohesion. In Product B, a highly flagged production file (23\% friction) with complex logic and frequent changes was %also 
seen as important, even if some issues were considered less severe than the metric suggested. %Likewise, 
In Product C-v1, one flagged file (45\% friction) was recognized as difficult due to its size and nested conditionals and was seen as a reasonable refactoring candidate. These suggest that friction aligned most clearly with maintenance relevance when a file was both structurally problematic and central to everyday development work.

\paragraph{Alignment was weaker for isolated files, test files, and temporary activity spikes}
%ARDA
%%%%%A third finding is that high friction did not always imply high practical importance. In Product B, some flagged files were test files. Although developers acknowledged quality issues in these files, they considered them less relevant to their use case than production code with similar scores. In Product C, one high-friction file was treated as lower priority because it was relatively isolated and had limited coupling to the rest of the codebase. Developers noted that the same friction level in a more central component would likely have been evaluated differently. The sessions also showed that some files were flagged mainly because of recent bursts of activity rather than because they represented persistent maintenance bottlenecks. Thus, friction aligned less well with maintenance relevance when the flagged file was peripheral, belonged to test code, or reflected a temporary development spike. A further nuance concerns test files. In some cases, practitioners treated flagged test files as lower priority than application code, even when the underlying issues were acknowledged. We do not interpret this as evidence that friction is less meaningful for test code. Rather, it reflects a short-term prioritization choice: poorly maintainable test code may still increase long-term maintenance cost by slowing validation, reducing confidence in change, and making refactoring harder to perform safely~\cite{garousiSmellsSoftwareTest2018}.

High friction did not always imply high practical importance. In Product B, some flagged files were test files (e.g., 17–19\% friction); although developers acknowledged their quality issues, they considered them less relevant than production files with similar scores. In Product C, one high-friction file (33\% friction) was treated as lower priority because it was relatively isolated and weakly coupled to the rest of the codebase; developers noted that the same friction level in a more central component would likely have been judged differently. The sessions showed that some files were flagged because of recent bursts of activity rather than persistent maintenance bottlenecks. Thus, friction aligned less well with maintenance relevance when the flagged file was peripheral, belonged to test code, or reflected a temporary development spike. We do not interpret the lower priority given to test files as evidence that friction is less meaningful there; rather, it reflects a short-term prioritization choice, since poorly maintainable test code may still increase long-term maintenance cost by slowing validation, reducing confidence in change, and making refactoring harder to perform safely~\cite{garousiSmellsSoftwareTest2018}.

\paragraph{Version comparisons suggest that friction is sensitive to both refactoring and ongoing work}
%ARDA
%%%%%A fourth finding is that version-to-version comparisons reveal both the usefulness and the limits of friction as an alignment signal. In several cases, refactoring reduced friction and improved code health, supporting the view that the metric can track meaningful maintenance progress. At the same time, the comparisons also showed that friction may remain high, or even temporarily increase, when a file is actively being reworked. This was particularly visible in Product C-v1, where developers believed one file had been refactored during the observation window; friction decreased, but remained substantial because change activity stayed high. In Product A-v2, some files from the earlier snapshot remained in comparable condition, indicating persistent relevance, while others changed for reasons more closely tied to ongoing work than to stable architectural problems. These observations show that friction can reflect genuine maintenance relevance over time, but that it is also sensitive to the timing of active development.

Comparing versions revealed both the usefulness and the limits of friction as an alignment signal. In several cases, refactoring reduced friction and improved code health (e.g., from 22\% to 8\% in Product A), showing that the metric can track meaningful maintenance progress. At the same time, friction could remain high, or even increase temporarily, when a file was actively being reworked. This was particularly visible in Product C-v1, where one file (initially around 33\% friction) appeared to have been refactored during the observation window: friction decreased, but remained substantial because change activity stayed high. In Product A-v2, some previously flagged files remained in similar condition (e.g., low but stable friction around 3–4\%), indicating persistent relevance, while others changed in ways more closely tied to ongoing work than to stable architectural problems. These observations show that friction can reflect genuine maintenance relevance over time, but is also sensitive to the timing of active development.

\paragraph{Context determined whether alignment translated into practical importance}
%ARDA
%%%Participants consistently interpreted alignment through product context rather than through metric values alone. Developers considered whether a file was part of production or test code, whether it was central or isolated, whether it was currently under refactoring, and whether recent commits reflected ordinary maintenance or temporary restructuring. This was especially clear in Product C, where the system was largely in maintenance mode and some files showed friction mainly because of concentrated work by a small number of contributors. Developers emphasized that such files could still be less urgent than centrally used components with similar scores. More generally, these discussions indicate that alignment with friction was not sufficient by itself; the practical significance of that alignment depended on contextual knowledge of the file's role and the broader development situation.
Participants consistently interpreted alignment through product context rather than metric values alone. They considered whether a file belonged to production or test code, whether it was central or isolated, whether it was under refactoring, and whether recent commits reflected routine maintenance or temporary restructuring. This was %especially 
clear in Product C, where the system was largely in maintenance mode and some files showed friction (e.g., 33–45\%) mainly because of concentrated work by a small number of contributors. Participants emphasized that such files could still be less urgent than centrally used components with similar scores. These discussions indicate that alignment with friction was not sufficient by itself; its practical significance depended on contextual knowledge of the file’s role and the broader development situation.

\begin{tcolorbox}[boxsep=2pt,left=2pt,right=2pt,top=2pt,bottom=2pt]

\textbf{RQ2 Conclusion.} Our findings show that \emph{Technical Debt Friction} aligned well with many areas that later proved relevant to maintenance and improvement, particularly central production files with persistent structural issues. However, the alignment was not uniform. It weakened for isolated files, test code, and files affected by temporary activity spikes, and it remained sensitive to ongoing refactoring work. These results suggest that friction is a useful indicator of maintenance-relevant areas, but that its practical value depends on reading it together with architectural role, file type, and current development context.
\end{tcolorbox}

%\subsection{RQ3: Benefits, Limitations, and Likely Challenges for Practical Use (Markus?)}

\subsection{RQ3: Broader Maintenance and Evolution Insights Beyond Individual Refactoring Candidates}
\label{sec:res_rq3}

%ARDA
%%%%%To respond to \textit{RQ3}, we complemented the file-level analyses from RQ1 and RQ2 with an exploratory project-level analysis of friction distributions. Rather than focusing on individual flagged files, we examined how touches were distributed across all files in the latest available versions of the studied products and related these activity patterns to code health. Figures~\ref{fig:td_friction_overview_a}--\ref{fig:td_friction_overview_c} visualize these distributions for Products A-v2, B, and C-v1, respectively. Our goal was to investigate whether project-level friction distributions can reveal broader maintenance and evolution characteristics beyond identifying individual refactoring candidates.

To address \textit{RQ3}, we complemented the file-level analyses from RQ1 and RQ2 with an exploratory project-level analysis of friction distributions. Instead of focusing on individual flagged files, we examined how touches were distributed across all files in the latest available versions of the studied products and related these patterns to code health. Our goal was to investigate whether project-level friction distributions can reveal broader maintenance and evolution characteristics beyond individual refactoring candidates. 

\begin{figure}[t]
    \centering
    \includegraphics[width=\linewidth]{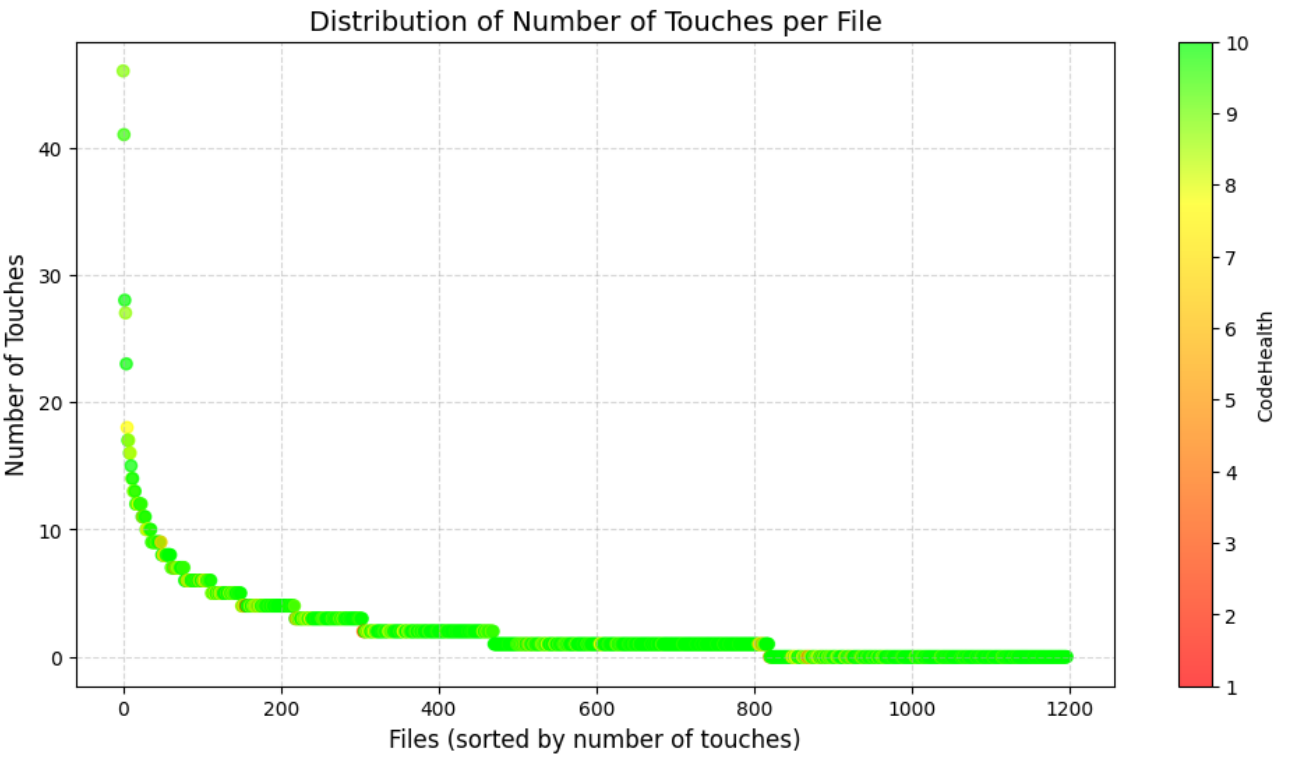}

    \vspace*{-0.8em}

    %%%\caption{Overall project friction for Project A~v2}
    \caption{Project-level friction distribution for Product A-v2. It plots the number of touches per file during the analysis period, with files ordered by touch frequency and colored by Code Health.} %%%The distribution indicates activity spread across many files, while also showing that part of this activity remains concentrated in files with lower code health.}
    \label{fig:td_friction_overview_a}

    \vspace*{-1.8em}

\end{figure}

\begin{figure}[t]
    \centering
    \includegraphics[width=\linewidth]{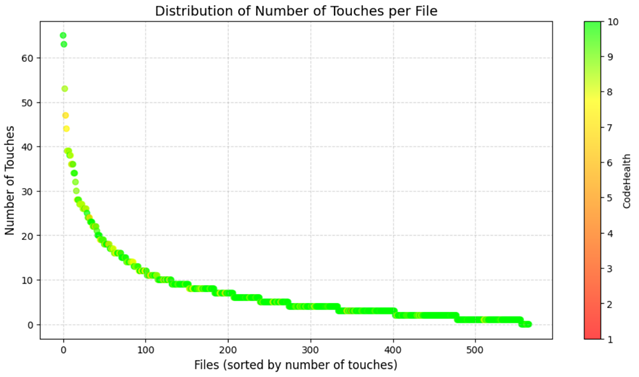}

    \vspace*{-0.8em}

    %%%\caption{Overall project friction for Project B}
    %%%%%\caption{Project-level friction distribution for Product B. The figure shows a broad spread of touched files, with most active files remaining relatively healthy. Compared with the other products, this profile suggests a more widely distributed development pattern with less concentration of activity in structurally weak code.}
    \caption{Project-level friction distribution for Product B. It shows a broad spread of touched files, with most active files remaining relatively healthy.} %Compared with the other products, this profile suggests a more widely distributed development pattern with less concentration of activity in structurally weak code.}
    \label{fig:td_friction_overview_b}

    \vspace*{-1.4em}

\end{figure}

\begin{figure}[t]
    \centering
    \includegraphics[width=\linewidth]{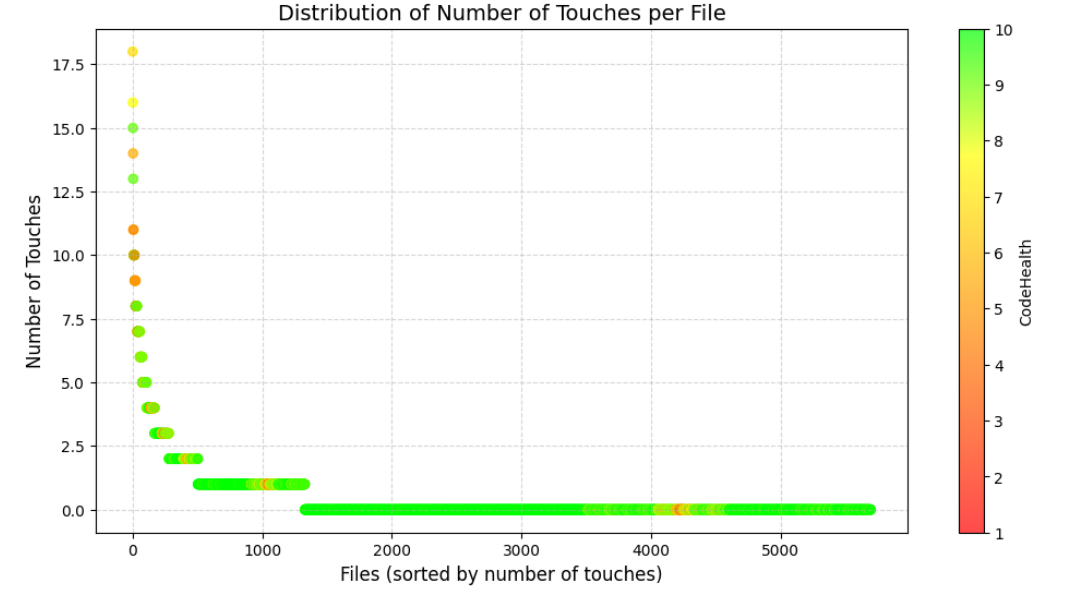}

    \vspace*{-0.8em}

    %\caption{Overall project friction for Project C~v2}
    \caption{Project-level friction distribution for Product C-v2. It shows that most files receive little or no activity, while a smaller subset accounts for the visible maintenance effort.} %This concentrated profile is consistent with a product in a more limited maintenance mode, where work is focused on selected parts of the codebase.}
    \label{fig:td_friction_overview_c}

    \vspace*{-1.8em}

\end{figure}

Figures~\ref{fig:td_friction_overview_a}--\ref{fig:td_friction_overview_c} show the distributions for Products A-v2, B, and C-v2, respectively. In each plot, files are ordered along the x-axis by decreasing number of touches (revisions) during the analysis period, while the y-axis shows the corresponding number of touches, revealing the distribution of recent development activity across the codebase. Each point represents a file, colored by its CodeHealth score on the 1--10 scale from red (low) to green (high). The plots thus combine the two main constituents of friction: change frequency and maintainability. 

%ARDA
%%%%Overall, the three products exhibit clearly different distribution shapes. This indicates that project-level friction does not merely summarize the presence of problematic files, but can also reveal how maintenance and development effort is distributed across the codebase. In particular, the figures suggest differences in activity concentration, spread of touched files, and the extent to which active files remain structurally healthy. At the same time, we treat these observations as exploratory: the current evidence supports qualitative contrasts across the studied products, whereas stronger claims about generality or classification require further quantitative validation.

The three products exhibit clearly different distribution shapes, suggesting that project-level friction captures more than the presence of problematic files. It also reveals how maintenance and development effort is distributed across the codebase, including differences in activity concentration, spread of touched files, and the extent to which active files remain structurally healthy. At the same time, these observations remain exploratory: the current evidence supports qualitative contrasts across the studied products, while stronger claims about generality or classification require further %quantitative 
validation.

\paragraph{The distributions reveal different maintenance and evolution profiles}
%ARDA
%%%%A first finding is that the overall distribution shapes differ markedly across the studied products. In Figure~\ref{fig:td_friction_overview_a}, Product A-v2 shows a broad distribution with visible activity across many files, but also a noticeable concentration of files with lower Code Health among the more frequently touched parts of the system. In Figure~\ref{fig:td_friction_overview_b}, Product~B displays an even broader spread of touched files, but with most of the visible points remaining green, indicating comparatively healthy code despite substantial activity. In contrast, Figure~\ref{fig:td_friction_overview_c} shows that Product C-v1 has a much more concentrated profile: most files receive little or no activity, while a smaller subset accounts for the visible maintenance effort. Taken together, these differences suggest that project-level friction distributions can capture broader product-level maintenance profiles rather than only local refactoring needs.

The overall distribution shapes differ markedly across the studied products. In Figure~\ref{fig:td_friction_overview_a}, Product A-v2 shows a moderately concentrated distribution, with a small set of highly active files and a long tail of rarely touched ones. Most points are green, indicating generally healthy active code, though a few lower-CodeHealth files appear among the more frequently touched files. %with a noticeable concentration of lower Code Health among the more frequently touched ones. 
In Figure~\ref{fig:td_friction_overview_b}, Product B has an even broader spread of touched files, with most points remaining green (comparatively healthy code despite substantial activity). By contrast, Figure~\ref{fig:td_friction_overview_c} shows that Product C-v2 has a much more concentrated profile, with most files receiving little or no activity and a smaller subset accounting for the visible maintenance effort. These differences suggest that project-level friction distributions can capture broader maintenance profiles rather than only local refactoring needs.

\paragraph{Project-level friction scores complement the visual profiles}
The visual contrasts in Figures~\ref{fig:td_friction_overview_a}--\ref{fig:td_friction_overview_c} are also reflected in the aggregate friction scores. Product A-v2 has a Dynamic Friction of 6.39 %, a TD Start Friction of 1.10, 
and a Total Friction of 7.49, compared with 7.69 %, 1.80, 
and 9.49 for Product A-v1. This reduction is consistent with a less severe overall friction profile in the newer version. Product B has values of 7.76 %, 1.40, 
and 9.16, respectively. %While its total friction is higher than that of A-v2, 
Figure~\ref{fig:td_friction_overview_b} shows that much of the activity occurs in comparatively healthy files. Product C-v2 has the highest scores by far---14.87 %, 3.70, 
and 19.27 (from 12.71 and 16.41 for Product C-v1)---which aligns with Figure~\ref{fig:td_friction_overview_c}, where a smaller subset of files concentrates much of the visible maintenance effort and lower CodeHealth values are more prominent. The friction scores reinforce the qualitative interpretation of the figures by showing clear differences in the magnitude of friction burden across the studied products.

\paragraph{The figures suggest different degrees of activity concentration}
%ARDA
%%%%%A second finding is that the figures make visible how strongly development and maintenance work is concentrated. In Product C-v1 (Figure~\ref{fig:td_friction_overview_c}), the distribution is dominated by a long flat region close to the x-axis, indicating that most files are touched rarely, while only a limited number of files receive notable attention. This pattern is consistent with a product in a more maintenance-oriented mode, where activity is concentrated in selected parts of the codebase. By contrast, Product~B (Figure~\ref{fig:td_friction_overview_b}) shows a much broader spread of files with non-trivial activity levels, suggesting a more actively evolving product in which development effort is distributed across a larger portion of the system. Product A-v2 (Figure~\ref{fig:td_friction_overview_a}) appears to lie between these two cases, combining broader activity than Product~C with more visible concentration of lower-health files than Product~B. These contrasts suggest that project-level friction distributions may help distinguish between products with concentrated maintenance effort and those with more broadly distributed evolution activity.

%The figures make visible how strongly development and maintenance work is concentrated. 
In Product C-v2 (Figure~\ref{fig:td_friction_overview_c}), a long flat region near the x-axis shows that most files are touched rarely, while only a limited subset receives notable attention. %, consistent with a more maintenance-oriented mode. 
This pattern is consistent with a product in a more maintenance-oriented mode, where activity is concentrated in selected parts of the codebase. By contrast, Product B (Figure~\ref{fig:td_friction_overview_b}) shows a much broader spread of files with non-trivial activity, suggesting more widely distributed evolution work. Product A-v2 (Figure~\ref{fig:td_friction_overview_a}) lies between these two cases, with broader activity than Product C but a more concentrated profile than Product B, while remaining largely healthy overall. These contrasts suggest that project-level friction distributions may help distinguish between concentrated maintenance effort and more broadly distributed development activity.
%ARDA
%%%%%Product A-v2 (Figure~\ref{fig:td_friction_overview_a}) lies between these two cases, with broader activity than Product C but more concentration of lower-health files than Product B. These contrasts suggest that project-level friction distributions may help distinguish between concentrated maintenance effort and more broadly distributed development activity.

% NOTE TO US:
% If we later want to summarize these differences quantitatively, this is a natural place to introduce a distribution summary measure such as AUC. To support that claim, we should define the curve formally, report AUC values for all product versions, and justify why AUC captures the relevant differences in spread or concentration.

\paragraph{Code Health adds interpretive value beyond activity alone}

The distributions become more informative when activity is interpreted together with Code Health. Because the figures encode Code Health through color, they show not only which files are active, but also whether that activity is concentrated in healthier or more problematic parts of the codebase. This is especially useful when comparing Products A-v2 and B: both show broad activity, but Figure~\ref{fig:td_friction_overview_a} suggests a somewhat more concentrated activity profile for A-v2, while Figure~\ref{fig:td_friction_overview_b} is more broadly distributed and remains dominated by green points. This points to two different maintenance situations: one in which activity is more concentrated, though still largely healthy overall, and another in which activity is more broadly distributed while the code remains healthy. Product C-v2 (Figure~\ref{fig:td_friction_overview_c}) adds that limited overall activity can still coexist with a smaller set of files that continue to attract maintenance effort. Thus, the project-level view is useful because it combines where activity occurs with an indication of the quality of the files involved.

\paragraph{Project-level friction broadens the analytical role of friction}
%ARDA
%%%%%A fourth finding is that the project-level perspective provides insights beyond identifying individual refactoring candidates. At the file level, friction is primarily useful for locating specific candidates for inspection and possible refactoring. At the project level, however, the distributions in Figures~\ref{fig:td_friction_overview_a}--\ref{fig:td_friction_overview_c} make visible broader properties of the maintenance landscape: whether work is concentrated or widespread, whether frequent changes accumulate in comparatively healthy or weaker code, and whether the overall profile is more consistent with active evolution or selective maintenance. In this sense, project-level friction broadens the analytical role of friction from local prioritization to higher-level characterization of how maintenance effort is distributed across a product.

The project-level perspective provides insights beyond individual refactoring candidates. While file-level friction is mainly useful for locating specific inspection or refactoring targets, the distributions in Figures~\ref{fig:td_friction_overview_a}--\ref{fig:td_friction_overview_c} reveal broader properties of the maintenance landscape, such as whether work is concentrated or widespread, whether frequent changes accumulate in healthier or weaker code, and whether the overall profile is more consistent with active evolution or selective maintenance. Project-level friction broadens the %analytical 
role of friction from local prioritization to higher-level characterization of how maintenance effort is distributed across a product.

\paragraph{The observed differences are promising, but the claims should remain bounded}
%ARDA
%%%%%Although the contrasts between Figures~\ref{fig:td_friction_overview_a}--\ref{fig:td_friction_overview_c} are clear, the current evidence supports only bounded conclusions. The figures show that the studied products exhibit different friction-distribution shapes and that these shapes align plausibly with what is known about their development context. However, stronger claims---for example, that such distributions can reliably distinguish lifecycle stage, maintenance mode, or project class---require additional support. Any such claim should be backed by a formal summary measure, systematic comparison across products or versions, and a broader evaluation set.

Although the contrasts between Figures~\ref{fig:td_friction_overview_a}--\ref{fig:td_friction_overview_c} are clear, the current evidence supports only bounded conclusions. The figures show that the studied products exhibit different friction-distribution shapes that plausibly align with their development contexts. However, stronger claims---such as reliably distinguishing lifecycle stage, maintenance mode, or project class---require additional support through a formal summary measure, systematic comparison across products or versions, and a broader evaluation set.

% NOTE TO US:
% If we want to retain lifecycle-related interpretations, we should support them with explicit contextual evidence for each product (e.g., maintenance mode, active expansion, brownfield evolution) and phrase them as "consistent with" or "suggestive of" unless independently validated.

\begin{tcolorbox}[boxsep=2pt,left=2pt,right=2pt,top=2pt,bottom=2pt]
\textbf{RQ3 Conclusion.} %Our exploratory analysis suggests that 
Project-level friction distributions provide broader maintenance and evolution insights beyond individual refactoring candidates. Figures~\ref{fig:td_friction_overview_a}--\ref{fig:td_friction_overview_c} and the corresponding aggregate scores reveal clear differences across products in activity spread, touch concentration, the relation between activity and CodeHealth, and overall friction magnitude. These results suggest that project-level views complement file-level analysis by characterizing broader maintenance profiles. However, the findings remain exploratory; stronger claims about lifecycle characterization or comparative ranking require further quantitative validation.

\end{tcolorbox}

\section{Cross-case Discussion}
\label{sec:discussion}

\begin{table*}[t]
\centering
\caption{Cross-case synthesis of how practitioners engaged with Technical Debt Friction across the study products. The table summarizes recurring interpretation patterns observed in the sessions and their implications for understanding friction as a maintenance-oriented concept.}
\label{tab:crosscase_friction_interpretation}
\vspace*{-1.0em}
\scriptsize
\begin{tabular}{p{2.95cm} p{5.7cm} p{0.9cm} p{1.5cm} p{5.1cm}}
\toprule
\textbf{Cross-case pattern} & \textbf{Illustrative evidence from sessions} & \textbf{Products} & \textbf{Analysis level} & \textbf{Discussion implication} \\
\midrule

\rowcolor{black!6}
Friction was grounded in concrete maintenance locations
& Repeated discussion of named subsystems, flagged files, and concrete code artifacts rather than abstract scores
& A, B, C
& Subsystem, file
& Friction was meaningful to practitioners when tied to real maintenance locations in the codebase \\

Friction was interpreted through neighboring views
& Code Health, Hotspots, Friction, Change Coupling, and Refactoring Targets were repeatedly used together
& A, B, C
& Cross-view
& Friction was treated as broader than any single metric and interpreted through the relation between structural quality, activity, and dependencies \\

\rowcolor{black!6}
Friction remained meaningful across levels of granularity
& Sessions moved between system maps, file-level examples, and function-oriented inspection
& A, B
& Subsystem, file, function
& Friction supported reasoning across both broader problematic areas and specific code artifacts \\

Friction included socio-technical concerns
& Ownership- and knowledge-related categories such as knowledge islands and contributor concentration were part of the discussion
& A
& Team, file
& Practitioners did not treat friction as purely structural, but also as shaped by knowledge distribution and coordination \\

\rowcolor{black!6}
Friction was treated as cumulative maintenance burden
& Files were discussed in relation to low code health, frequent change, coupling, complexity, and review-related issues
& A, B, C
& File, function
& Friction was understood as the combined effect of several maintenance-relevant conditions rather than a single defect type \\

Friction was discussed in a prioritization-oriented way
& Friction was repeatedly considered in relation to whether files warranted further attention or refactoring discussion
& A, B, C
& Cross-view, file
& Friction supported reasoning about where maintenance effort and improvement attention might be justified \\

\rowcolor{black!6}
Interpretation depended on explanatory context
& Friction was consistently discussed together with supporting information such as code health values, hotspot context, coupling, and team-related views
& A, B, C
& Cross-view
& Friction was useful in context, but not treated as self-explanatory in isolation \\

Friction should be read as a context-sensitive indicator
& Across products, practitioners qualified flagged files using architectural role, file type, coupling, maintenance mode, and knowledge of ongoing work
& A, B, C
& File, product
& Friction is most useful when interpreted together with product context rather than as a universal signal of importance \\

\rowcolor{black!6}
Alignment did not eliminate the need for judgment
& Even when flagged files appeared meaningful, practitioners still differentiated between files worth immediate attention and files better treated as lower priority
& A, B, C
& File, cross-case
& Friction can support prioritization, but does not replace human assessment of practical importance \\

Temporal interpretation matters
& Across versions and follow-up discussions, practitioners considered whether friction reflected accumulated burden, recent spikes in activity, or ongoing rework
& A, C
& File, version
& Friction should be interpreted in light of development timing, since high values may reflect either persistent problems or active remediation \\

\rowcolor{black!6}
Friction is more useful for structured discussion than for automatic ranking
& Across cases, friction was most effective when used to guide inspection and discussion, not when treated as a definitive ordering of files
& A, B, C
& Cross-case
& The main value of friction lies in supporting maintenance reasoning and communication rather than fully automating prioritization \\

Project-level friction broadens the analytical role of friction
& Beyond individual files, the product-level distributions helped characterize how maintenance effort was distributed across the codebase as a whole
& A, B, C
& Product, version
& Friction can support higher-level reasoning about maintenance and evolution, not only local refactoring candidate identification \\

\rowcolor{black!6}
Distribution shape can reveal different maintenance profiles
& Across products, the observed distributions differed in how strongly touches were concentrated in a limited subset of files versus spread across a larger part of the codebase
& A, B, C
& Product, version
& Project-level friction may help distinguish between more concentrated maintenance effort and more broadly distributed development activity \\

Activity alone is insufficient without code health context
& The project-level plots became more informative when touch distributions were interpreted together with Code Health, revealing whether active parts of the system were predominantly healthy or problematic
& A, B, C
& Product, version
& The broader value of friction lies in combining activity patterns with structural quality rather than inspecting either dimension alone \\

\rowcolor{black!6}
Project-level friction is promising, but its broader use remains exploratory
& The product-level analysis revealed meaningful contrasts across the studied products, but stronger claims about lifecycle characterization or comparative ranking would require additional quantitative validation
& A, B, C
& Product, version
& Project-level friction distributions are useful as exploratory maintenance indicators, but should not yet be treated as validated classification or benchmarking measures \\

\bottomrule
\end{tabular}
\vspace*{-1.2em}
\end{table*}

%ARDA
%%%%The findings of the RQs show that \emph{Technical Debt Friction} is best understood not as a standalone metric, but as a maintenance-oriented decision-support concept whose usefulness depends on context, supporting evidence, and practitioner interpretation. Table~\ref{tab:crosscase_friction_interpretation} summarizes the main cross-case patterns that emerged across the three products. Rather than repeating the case-level results, this section discusses what those combined findings imply for how friction should be interpreted and used in practice.

The RQ findings show that \emph{Technical Debt Friction} is best understood not as a standalone metric, but as a maintenance-oriented decision-support concept whose usefulness depends on context, supporting evidence, and practitioner interpretation. 
%ARDA
%%%Table~\ref{tab:crosscase_friction_interpretation} summarizes the main cross-case patterns across the three products. Rather than repeating the case-level results, this section discusses what those combined findings imply for how friction should be interpreted and used in practice.
Table~\ref{tab:crosscase_friction_interpretation} presents the detailed cross-case patterns observed across the products. In the discussion below, we synthesize these patterns into five higher-level cross-case insights (Sections~\ref{subsec:integrative-concept}--\ref{subsec:usefulness-discussion}).  Rather than repeating the RQ results, this section discusses what the combined findings imply for how \emph{Technical Debt Friction} should be interpreted and used in practice across products, contexts, and levels of analysis.

%%%More specifically, the first insight draws on the patterns related to concrete maintenance locations, neighboring views, levels of granularity, socio-technical concerns, cumulative maintenance burden, and explanatory context. The second builds on the patterns concerning context sensitivity and the need for judgment. The third focuses on the role of friction in bridging diagnosis and prioritization. The fourth synthesizes the temporal patterns observed across versions and ongoing work. The fifth extends the discussion to the project-level patterns related to friction distributions, activity concentration, code health context, and the exploratory role of project-level friction.

\subsection{Friction as an Integrative and Context-Sensitive Concept}
\label{subsec:integrative-concept}
%ARDA
%%%%%A first cross-case insight is that practitioners engaged with friction as an \emph{integrative} concept. Across all products, friction became meaningful when discussed together with code health, hotspots, change coupling, refactoring-oriented views, and, where available, socio-technical information. Its value, therefore, lay less in introducing a fundamentally new signal than in bringing together several maintenance-relevant signals into a single maintenance-oriented perspective.

Practitioners engaged with friction as an \emph{integrative} concept. Across all products, it became meaningful when discussed together with code health, hotspots, change coupling, refactoring-oriented views, and, where available, socio-technical information. Its value therefore lies less in introducing a new signal than in combining several maintenance-relevant signals into a single maintenance-oriented perspective.

At the same time, friction is inherently \emph{context-sensitive}. Its meaning depends on factors such as file role, architectural centrality, coupling, maintenance mode, and ongoing work. Practitioners did not treat flagged files as important based on friction alone, but interpreted them in context (e.g., production vs. test, central vs. isolated, persistent vs. temporary activity). Thus, friction is most useful when supporting contextual interpretation rather than as a universal signal of importance.

%ARDA
%%%A particularly important example is the recurring distinction between test and application code. Practitioners sometimes treated friction in test files as less urgent, even when those files were visibly problematic. This should not be read as evidence that the signal is less meaningful for the test code. Rather, it reflects a prioritization pattern in which immediate delivery concerns outweigh longer-term maintainability considerations. In practice, however, poorly maintainable test code can still slow validation, complicate regression analysis, reduce confidence in change, and increase the cost of future refactoring. Discounting test-code friction may therefore reflect a hidden maintenance burden rather than a flaw in the signal.

A key example is the distinction between test and application code. Practitioners sometimes treated friction in test files as less urgent, even when problematic. This does not mean the signal is less meaningful; rather, it reflects short-term prioritization where delivery concerns outweigh long-term maintainability. In practice, poorly maintainable test code can still slow validation, complicate regression analysis, reduce confidence in change, and increase future refactoring costs. Discounting test-code friction may therefore indicate hidden maintenance burden rather than a flaw in the signal.

\subsection{Friction Bridges Diagnosis and Prioritization}
\label{subsec:bridges-diagnosis}
%ARDA
%%%%%A second cross-case insight is that friction helps bridge the gap between \emph{diagnosis} and \emph{prioritization}. Existing software analytics often provide valuable but fragmented views of structural weaknesses, development activity, coupling, or team-related risks. Across the cases, practitioners used friction to connect these views to the more practical question of where attention might be warranted. This does not mean that friction replaced prioritization judgment. Practitioners still distinguished between files that were immediately relevant, files that were plausible but lower priority, and files whose scores reflected temporary or peripheral conditions. However, the recurring discussion pattern suggests that friction was useful in focusing attention and structuring conversation around whether the improvement effort might be justified. In this sense, friction narrowed the search space of concern rather than automating the final decision.

Friction helps bridge the gap between \emph{diagnosis} and \emph{prioritization}. Existing software analytics often provide valuable but fragmented views of structural weaknesses, development activity, coupling, or team-related risks. Across the cases, practitioners used friction to connect these views to the more practical question of where attention might be warranted. This did not replace prioritization judgment: practitioners still distinguished between immediately relevant files, plausible but lower-priority files, and files whose scores reflected temporary or peripheral conditions. However, friction helped focus attention and structure the discussion around whether the improvement effort might be justified. It narrowed the search space of concern rather than automating the final decision.

\subsection{The Importance of Time and Ongoing Activity}
\label{subsec:importance-time}
%ARDA
%%%%%A third cross-case insight is that friction must be interpreted in relation to \emph{development timing}. The version comparisons and follow-up observations showed that friction may reflect both persistent maintenance burden and active remediation. In some cases, refactoring reduced friction and improved code health, while in others, friction remained elevated during active cleanup because changes continued while quality improved more slowly. This means that high friction can point either to accumulated burden or to ongoing work intended to reduce that burden. This temporal sensitivity reinforces that friction should not be read statically or independently of current development activity. It appears most informative when interpreted together with knowledge of whether a component is being actively reworked, stabilized, or only occasionally modified.

%A third cross-case insight is that 
Friction must be interpreted in relation to \emph{development timing}. The version comparisons and follow-up observations showed that friction may reflect both persistent maintenance burden and active remediation. In some cases, refactoring reduced friction and improved code health; in others, friction remained elevated during active cleanup because changes continued while quality improved more slowly. High friction may therefore indicate either accumulated burden or ongoing work intended to reduce it. This temporal sensitivity reinforces that friction should not be read statically or independently of current development activity, but interpreted %together 
with knowledge of whether a component is being actively reworked, stabilized, or only occasionally modified.

\subsection{Project-Level Characterization}
\label{subsec:project-level}

The project-level analysis suggests that friction can support more than identifying individual refactoring candidates by showing how maintenance effort is distributed across a product. The studied products exhibited different distribution shapes, indicating varying degrees of activity concentration versus spread across the codebase, which may help distinguish between concentrated maintenance effort and more distributed development activity. The plots also show that activity is more informative when interpreted with code health, revealing whether work accumulates in healthier or more problematic parts of the system. At the same time, this broader use remains exploratory: although the contrasts were meaningful and consistent with development contexts, stronger claims about lifecycle characterization or comparative ranking require further quantitative validation. Project-level friction should therefore be seen as a promising exploratory indicator rather than a validated benchmarking measure.

\subsection{Usefulness for Discussion, Not Automatic Ranking}
\label{subsec:usefulness-discussion}
%ARDA
%%%%%A final cross-case insight is that friction appears more useful for \emph{structured maintenance discussion} than for \emph{automatic ranking}. Across products, practitioners relied on friction as a starting point for inspection and reflection, but not as a self-sufficient answer to what should be done next. Even when flagged files aligned well with known problematic areas, contextual knowledge and architectural understanding remained necessary before deciding whether a file truly deserved refactoring or maintenance attention.

%A final cross-case insight is that 
Friction appears more useful for \emph{structured maintenance discussion} than for \emph{automatic ranking}. Across products, practitioners used it as a starting point for inspection and reflection, not as a self-sufficient answer to what should be done next. Even when flagged files aligned with known problematic areas, contextual and architectural understanding remained necessary before deciding whether a file truly warranted refactoring or maintenance attention.

\subsection{Implications}
\label{subsec:implications-insights}

These findings suggest three implications. First, \emph{Technical Debt Friction} should be treated as a context-sensitive concept for reasoning about accumulated maintenance burden, not as a universally interpretable score. Second, its practical value lies in connecting fragmented software analytics to maintenance prioritization and communication. Third, effective use of friction requires supporting evidence and human judgment, especially when flagged files differ in role, centrality, or development timing. Overall, the combined findings indicate that friction is promising as an industrial maintenance concept because it helps practitioners reason not only about where code is problematic, but also about where the system resists change in ways that matter for development work.

\section{Threats to Validity}
\label{sec:empirical:validity} 

\textbf{Construct validity.}
%%%%%A key threat concerns the interpretation of \emph{Technical Debt Friction}, which combines technical and socio-technical aspects of maintenance burden and may therefore be understood differently by different practitioners. In addition, the study relied on CodeScene's operationalization of friction and related views. We mitigated this threat by grounding the discussions in concrete system artifacts and by examining friction together with complementary analyses rather than in isolation.
A key threat is the interpretation of \emph{Technical Debt Friction}, which combines technical and socio-technical aspects and may be understood differently by practitioners. The study also relies on CodeScene’s operationalization of friction and related views. We mitigated this by grounding discussions in concrete artifacts and interpreting friction with complementary analyses rather than in isolation.

\textbf{Internal validity.}
%%%The sessions may have been influenced by interviewer framing, the order of presented views, or participants' prior beliefs about the products. Because the study was walkthrough-based, we cannot exclude confirmation bias or selective attention to familiar examples. We reduced this risk through a shared interview protocol, multiple cases, and discussion of both confirming and disconfirming examples.
The sessions may have been influenced by interviewer framing, presentation order, or participants’ prior beliefs. As a walkthrough-based study, confirmation bias and selective attention cannot be excluded. We mitigated this through a shared interview protocol, multiple cases, and discussion of both confirming and disconfirming examples.

\textbf{External validity.}
%%%The study covers three industrial products, two with multiple versions, and may therefore not generalize to all systems, domains, or organizational contexts. The usefulness of friction may depend on factors such as system size, architecture, maintenance mode, team structure, and repository history. We therefore interpret the findings as evidence of practical relevance in the studied settings rather than universally generalizable conclusions.
The study covers three industrial products (two with multiple versions) and may not generalize to all systems, domains, or organizations. The usefulness of friction may depend on factors such as system size, architecture, maintenance mode, and team structure. %, and repository history. 
We interpret the findings as evidence of practical relevance in the studied settings rather than universally generalizable conclusions.

\textbf{Reliability.}
%%%As in many exploratory industrial studies, the analysis involves researcher interpretation of practitioner discussions and observations. We mitigated this threat by structuring the study around explicit research questions, using a shared interview protocol, and grounding the findings in concrete examples and follow-up comparisons across versions. Nevertheless, some degree of subjectivity remains unavoidable.
As in many exploratory industrial studies, the analysis involves researcher interpretation of practitioner discussions. We mitigated this through explicit research questions, a shared interview protocol, and grounding findings in concrete examples and version comparisons. Nevertheless, some subjectivity remains unavoidable.

%%%%%\textbf{Exploratory project-level analysis.} The project-level friction analysis in RQ3 is more exploratory than the interview-based analyses in RQ1 and RQ2. While the observed distribution differences are informative, stronger claims about lifecycle characterization or comparative ranking require broader quantitative validation. We therefore present these results as exploratory evidence only.

%%%%%Overall, these threats do not invalidate the study, but they indicate that the findings should be interpreted as an initial empirical assessment of the practical usefulness and limitations of \emph{Technical Debt Friction}.

\section{Related Work}
\label{section:related-work}

%ARDA
%%%%%Our work relates to research on technical debt~\cite{kruchten2012technical, li2015systematic, allman2012managing}, maintenance prioritization~\cite{bennett2000software, lientz1980software}, refactoring support~\cite{mens2004survey, baqais2020automatic}, and socio-technical software evolution~\cite{mens2016ecosystemic, hoda2021socio}. Technical debt has long been used to describe how short-term technical compromises increase future maintenance cost and complicate software evolution~\cite{kruchten2012technical}. Subsequent work has argued that the practical challenge is not only to identify debt, but also to understand how it creates development \emph{friction} and affects productivity in real projects~\cite{avgeriou2015reducing}. Recent studies further show that technical debt prioritization depends on developer judgment, project context, and organizational trade-offs rather than on a single technical indicator alone~\cite{pina2022technical,fungprasertkul2024technical, alfayez2020systematic, lenarduzzi2021systematic, pina2021technical, de2019tracy, mensah2018value}. Our work builds on this line of research, but focuses specifically on \emph{technical debt friction} as a practice-oriented lens for identifying where technical debt most strongly affects ongoing maintenance and evolution.

Our work relates to research on technical debt~\cite{kruchten2012technical, li2015systematic, allman2012managing}, maintenance prioritization~\cite{bennett2000software, lientz1980software}, refactoring support~\cite{mens2004survey, baqais2020automatic}, and socio-technical software evolution~\cite{mens2016ecosystemic, hoda2021socio}. Technical debt describes how short-term compromises increase future maintenance cost and complicate evolution~\cite{kruchten2012technical}. Prior work shows that the challenge is not only to identify debt, but also to understand how it creates development \emph{friction} and affects productivity~\cite{avgeriou2015reducing}, and that prioritization depends on developer judgment, context, and organizational tradeoffs rather than a single technical indicator~\cite{pina2022technical,fungprasertkul2024technical, alfayez2020systematic, lenarduzzi2021systematic, pina2021technical, de2019tracy, mensah2018value}. Building on this, we focus on \emph{technical debt friction} as a practice-oriented lens for identifying where technical debt most strongly impacts ongoing maintenance and evolution.

A second line of work has shown that software analytics can support maintenance and refactoring decisions~\cite{noei2025empirical, chowdhury2022empirical}. Code churn and related evolutionary signals help identify change-prone components~\cite{nagappan2005use, chowdhury2025good}, while mining version histories reveals change coupling and hidden maintenance dependencies between artifacts~\cite{zimmermann2005mining, jin2020exploring}. Other work has combined code quality and development activity to identify parts of the system that deserve attention, which is closely related to hotspot-based reasoning~\cite{bohnet2011monitoring}. Refactoring studies likewise show that improvement decisions are selective and shaped by expected benefit, available tool support, and practical constraints rather than by structural quality considerations alone~\cite{kim2012field,murphy2008refactoring,ivers2022industry,nikolaidis2024metrics, murphy2011we, golubev2021one, tempero2017barriers}. Recent work on technical debt prioritization and automation also confirms continuing industrial interest in methods and tools that help rank debt items and guide action, while highlighting that actionability remains a central challenge~\cite{detofeno2022priortd,biazotto2024technical, pina2022technical, tsoukalas2024practical, biazotto2024technical, fungprasertkul2024technical}. In contrast to prior work that studies individual signals or automation mechanisms in isolation, our paper investigates whether practitioners find \emph{technical debt friction} meaningful and useful as an integrative decision-support concept across real industrial cases.

%ARDA
%%%%%Finally, maintenance burden is not purely technical. Ownership, expertise concentration, and coordination structure also influence software quality and productivity~\cite{bird2011don,cataldo2008socio, souza2024maintenance}. This socio-technical perspective is important because many maintenance problems emerge from the interplay of technical dependencies and organizational conditions rather than from isolated code properties alone~\cite{bird2011don,cataldo2008socio, lambiase2024cultural}. Our study complements this body of work by examining whether practitioners recognize technical debt friction as a way to interpret such combined effects when reasoning about maintenance pain points and prioritization.

Maintenance burden is not purely technical; ownership, expertise concentration, and coordination also influence software quality and productivity~\cite{bird2011don,cataldo2008socio, souza2024maintenance}. This socio-technical perspective matters because many maintenance problems arise from the interplay of technical dependencies and organizational conditions rather than isolated code properties~\cite{bird2011don,cataldo2008socio, lambiase2024cultural}. Our study complements this work by examining whether practitioners recognize \emph{technical debt friction} as a way to interpret these combined effects when reasoning about maintenance pain points and prioritization.

\section{Conclusion}
\label{sec:conclusion}

This paper presented a multi-case industrial study of \emph{technical debt friction} in real software development contexts. We examined how practitioners interpret the concept, how friction-related analyses align with maintenance-relevant areas, and whether project-level friction distributions can provide broader maintenance and evolution insights.

%ARDA
%%%%%Our findings show that practitioners generally regarded technical debt friction as a useful indicator of maintenance burden, but not necessarily as a standalone signal. At file level, friction often aligned with known problematic areas and, in several cases, with files that later received maintenance or refactoring attention. However, its practical relevance depended strongly on context, including file role, architectural centrality, and ongoing development activity. At project level, the exploratory analysis suggests that friction distributions may reveal broader maintenance and evolution patterns, although these observations require further quantitative validation.

Our findings show that practitioners generally regarded \emph{technical debt friction} as a useful indicator of maintenance burden, but not as a standalone signal. At file level, it often aligned with known problematic areas and, in several cases, with files that later received maintenance or refactoring attention, although its relevance depended strongly on context, such as file role, architectural centrality, and ongoing activity. At project level, the exploratory analysis suggests that friction distributions may reveal broader maintenance and evolution patterns, though these require further quantitative validation.

%%%%%Overall, the results indicate that \emph{technical debt friction} is promising as a maintenance-oriented decision-support concept. Its practical value lies in helping practitioners connect software analytics to maintenance prioritization, while its effective use depends on contextual interpretation, supporting evidence, and human judgment.

\section*{Acknowledgment} This paper has received funding from the Research Council of Norway’s Innovation Project for the Industrial Sector (IPN) Programme under grant agreement for project No. 340991 (TechDebtOps) and the Competence Centre NextG2Com funded by the VINNOVA program for Advanced Digitalisation with grant number 2023-00541.

%%%\newpage 

%\balance
\bibliographystyle{IEEEtran}
\bibliography{references}
\end{document}